\RequirePackage[2020-02-02]{latexrelease}
\documentclass[prb,twocolumn,showpacs,amsmath,amssymb]{revtex4} 
\usepackage{graphicx}
\usepackage{tabularx}
\usepackage{dcolumn}
\usepackage{bm} 
\usepackage{color} 
\usepackage{amsmath}

\begin{document}

\title{Multiple localized-itinerant dualities
in magnetism of 5f electron systems. 
The case of UPt$_2$Si$_2$}
 
\author{L. M. Sandratskii$^{1,2}$\footnote{lsandr3591@gmail.com}, V. M. Silkin$^{2,3,4}$, L. Havela$^5$}
\affiliation{$^1$Institute of Physics, Czech Academy of Sciences, 182 21 Prague, Czech Republic\\
$^2$Donostia International Physics Center (DIPC), Paseo de Manuel Lardizabal 4, E-20018 San Sebasti\'an, Spain\\
$^3$%Departamento de Física de Materiales, Facultad de Ciencias Quimicas, UPV/EHU, 20080 San Sebastian, Spain\\
Departamento de Pol\'{\i}meros y Materiales Avanzados: F\'{\i}sica,
Qu\'{\i}mica y Tecnolog\'{\i}a, Facultad de Ciencias
Qu\'{\i}micas, Universidad del Pa\'{\i}s Vasco (UPV-EHU), Apdo. 1072,
E-20080 San Sebasti\'an, Spain\\
$^4$IKERBASQUE, Basque Foundation for Science, 48011 Bilbao, Spain\\
$^5$Faculty of Mathematics and Physics, Charles University, 12116 Prague,
Czech Republic}

\begin{abstract}
The paper deals with the U based compound UPt$_2$Si$_2$ (UPS). The material was first treated 
as a localized 5f-electron system.
Later, an opposite opinion of a predominantly itinerant nature of the system was put forward.
The most recent publications treat UPS as a dual material. 
We suggest
a material specific theoretical model based on the density functional theory plus Hubbard $U$ (DFT+$U$) 
calculations that describes the set 
of fundamental ground-state properties and high magnetic field experiment. The ground state
properties include antiferromagnetic magnetic structure, magnetic
easy axis, and the value of the U atomic moment. The in-field experiment shows the presence of a 
strong metamagnetic transition for the field
parallel to the easy axis in contrast to the hard field direction where such a feature is absent.
On the other hand, comparable induced magnetization values are obtained for both easy and hard 
field directions. 
Within the framework of the suggested model we show that the compound possesses well-formed atomic
moments built by electrons treated as delocalized.
\textcolor{black}{
To understand the experimental high-field proprties we estimate exchange energy, magnetic anisotropy 
energy, and Zeeman energy. All three energies are shown
to have comparable values what is crucial for the interpretation of the experiment.
}
At all steps of the study we devote special attention to revealing
and emphasizing the dual itinerant-localized  
properties of the material. The obtained forms of
the duality are different: well defined atomic moments formed by the itinerant electrons,
interplay of the single-site and two-site anisotropies, strong localization of two of the 5f
electrons in contrast to the itinerant nature of the 5f electrons contributing to the states
around the Fermi level, intense Stoner continuum competing with spin wave formation. 
The obtained high sensitivity of the calculated properties to
the details of the theoretical model reflects the 
complexity of the multi-orbital 5f electron system. The latter is the origin of the 
wide range of complex behavior observed in U based materials.
\end{abstract}

\maketitle

\section{Introduction}
\label{sec_Intro}

The magnetic properties of the U-based compounds are highly diverse
(see, e.g., some reviews and recent publications 
\onlinecite{Sechovsky1998,Santini1999,Mydosh2017,IrkhinBook,
Miyake2018,Huston2022,Pikul2022,Isbill2022,Saniz2023,Ladia2023,Huxley2015,Mydosh2011}).
The magnetic structures of the materials vary widely including
ferromagnetism (FM), antiferromagnetism (AFM), spin-density waves (SDW), noncollinear commensurate 
and chiral incommensurate structures. 
Besides the rich variety of ground state properties
one finds numerous metamagnetic transitions (MMTs) whose features depend on the direction
of applied magnetic field. 
In spite of a vast amount of collected experimental and theoretical knowledge, the understanding 
of the U-based compounds is
far from complete. 

The complexity of the properties of the U based materials led to the suggestion
of a variety of physical models different
in their key assumptions. Some models treat the U 5f electrons and U magnetic moments as localized
and the physics of the materials as dominated
by the processes characteristic for isolated atoms.
To the localized-type properties belong, for example, the atomic magnetic moments formed
according to the Hund's rules
and many-electron atomic multiplets.
Other models emphasize the itinerant nature of the 5f magnetic moments
and the participation of the 5f orbitals in the formation of the Fermi surface (FS).
As can be expected in such a situation, there also exist 
dual models stressing coexistence of both localized atomic-like and 
collective itinerant features in the same U based material.

The complexity characteristic for the U based systems is clearly 
manifested in the case of UPt$_2$Si$_2$ (UPS), the compound 
that constitutes the topic of the present study.
In recent years, UPS attracted considerable research attention and was identified as the 
material with dual nature of magnetism. Looking at a longer time frame, 
we find that UPS was first treated as a
localized-electron magnet \cite{Amitsuka1992}. In later publications
it was suggested that the system is predominantly itinerant \cite{Elgazzar2012,Schulze2012}. 
In Ref.~\onlinecite{Schulze2017}, a dual theoretical treatment
of the 5f electrons was reported. The duality became the title property of
UPS in Ref.~\onlinecite{Lee2018} though the duality concept here is different from that
in Ref.~\onlinecite{Schulze2017}.
Also in the most recent publication \cite{Petkov2023} dealing with charge density wave, 
the material is characterized as dual magnet.

\begin{figure}[t]
\includegraphics*[width=6cm]{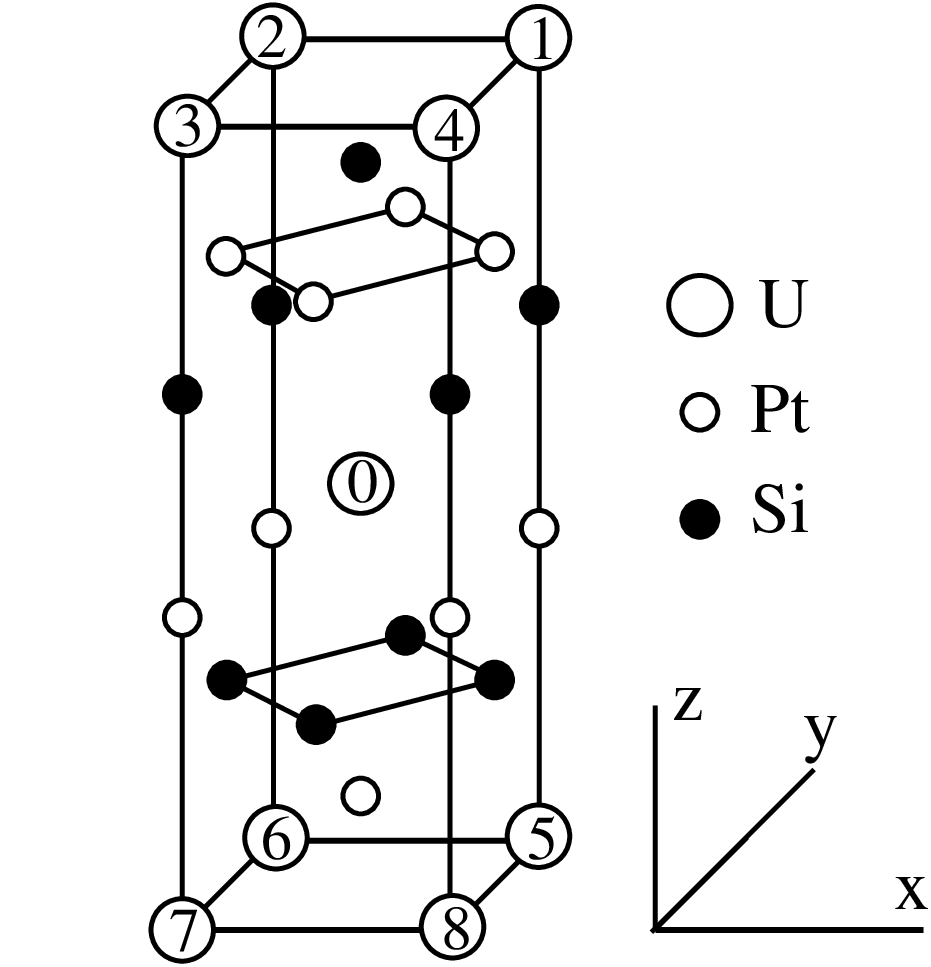}
\caption{Crystal structure of UPS. The $z$ axis corresponds to the crystallographic $c$ axis.
The $x$ axis corresponds to the crystallographic $a$ axis. The numbering of the U atoms is used 
in the symmetry analysis performed in Sec.~\ref{sec_calc_parameters}.
}
\label{Fig_lattice}
\end{figure}
The experimental information about fundamental magnetic properties of UPS is the following.
The material has ferromagnetic U layers that are antiferromagnetically coupled to each other
(Fig.~\ref{Fig_lattice}). The easy axis is orthogonal to the ferromagnetic layers. 
The large U moments up to 2.5 $\mu_B$ were reported\cite{Sullow2008}.
The high
magnetic field experiment \cite{Schulze2012} shows strong dependence of the magnetic response on the direction of applied 
field. For $B||z$ there is a distinct MMT while for $B||x$ such a clear feature is absent. 
At the same time, large induced magnetizations of comparable values 1.5 and 1.1 $\mu_B$/U 
are obtained in maximal field of $\sim$50~T 
for, respectively, easy and hard field directions. 
While early in-field experiments were interpreted
in terms of localized model, in Ref.~\onlinecite{Schulze2012} the itinerant model and major role of the Lifshitz 
transition were proposed\cite{Lifshitz1960,comment_Lifshitz}.
Interesting results of the inelastic neutron scattering (INS) measurements were reported 
in Ref.~\onlinecite{Lee2018}.
They have shown the presence of local U moments involved in transversal fluctuations
coexisting with the absence of any signature of spin wave excitations.

The material-specific density-functional-theory (DFT) based theoretical studies of 
the magnetic properties of UPS are scarce \cite{Elgazzar2012,Schulze2017}. 
In Ref.~\onlinecite{Elgazzar2012}, the authors report the comparison of the AFM and FM energies obtained 
in the local density approximation (LDA) that agrees with the experimental situation. 
The atomic U moment is calculated with three methods: LDA, 
LDA plus orbital polarization correction\cite{Eriksson1990} (LDA+OPC)  
and LDA+$U$ \cite{Anisimov1997}. 
As expected, the LDA result underestimates the value of the U atomic 
moment giving 0.71~$\mu_B$. This DFT weakness has been understood quite a time ago as the consequence of the 
underestimation of the orbital moment \cite{Eriksson1990}. 
The DFT+OPC method was suggested and intensively used 
to improve on this
issue. This method was directly designed to increase the value of the orbital moment (OM)
by introducing an effective orbital magnetic field \cite{Eriksson1990}.
In Ref.~\onlinecite{Elgazzar2012}, the use of the DFT+OPC method with a selected effective field parameter 
resulted in an increased total moment of 2.06~$\mu_B$.
The DFT+$U$
method has a more solid theoretical basis and has the ability, through introducing
orbital dependent potentials, to solve the problem of the OM
underestimation \cite{Solovyev1998}.
Rather unexpected, the application of the LDA+$U$ method in Ref.~\onlinecite{Elgazzar2012} resulted in a further 
reduction of the OM 
that became even smaller than the spin moment.
The FS was calculated and the 
conclusion about predominantly itinerant nature of the 5f electrons was drawn. 
In Ref.~\onlinecite{Schulze2017} the theoretical study of the FS was continued with application of 
the dual approach where two of the 5f electrons are treated as localized 
and the rest of the 5f electrons as itinerant.

The purpose of the present paper is a systematic material-specific study of UPS aiming at 
obtaining the theoretical model 
providing the agreement with experiment for the set of fundamental magnetic properties: 
the type of magnetic structure,
the character of magnetic anisotropy (MA), the value of atomic moments. The model thus gained is applied
to the interpretation of the high magnetic-field experiment. A possible microscopic reason of 
the absence of spin waves, as detected in the INS experiment, 
is also briefly addressed. 

Our study is based on the DFT and DFT+$U$ calculations for a number 
of magnetic configurations.
Special attention is given to the analysis of the calculation results  
from the viewpoint of the 
interplay of localized and delocalized properties of the 5f electron system.
In particular, we relate the physical picture following from our calculations
to the dual model of two separate groups
of the 5f electrons applied to UPS in Ref.~\onlinecite{Schulze2017}. 
For some $U_H$ values, we obtain and discuss the violation  
of the Bruno's relation \cite{Bruno1989,forBuno1,forBruno2}
between magnetic anisotropy energy (MAE) and 
OM anisotropy (OMA). 
(In the following we will use $U_H$ as the notation for the Hubbard parameter to easier 
distinguish it from the chemical symbol for uranium.)

The paper is structured as follows. In Sec.~\ref{sec_Duality} the relevant forms of
localized-itinerant dualities are discussed. Section~\ref{Sec_Method of calculations}
is devoted to the method of calculation. In Sec.~\ref{sec_Results and discussion} the 
results of the calculations are presented and discussed. 
The last section is devoted to the conclusions.

\section{Duality forms}
\label{sec_Duality}

The analysis of the experimental results and theoretical models in terms of 
the itinerant-localized duality helps to gain a deeper 
understanding of the magnetic systems. 
The concrete forms of duality can be 
very different. Therefore it is worthwhile to devote this short section to  
a brief preliminary discussion of the duality forms relevant to our study.

The dual behavior is obtained already on the level of individual electron states.
The DFT based treatment considers the 5f U orbitals as a part of the Bloch wave functions 
extended over whole crystal and, 
therefore, as a part of the delocalized electron picture. On the other hand, the spatial
distribution of the 5f electron density within the crystal volume assigned to the U atoms 
resembles the 5f density distribution in isolated U atoms. This duality in the properties 
of individual electron states is the origin of other duality types.

Historically, the understanding of the role of duality concept was crucial for solving the 
long-lasting conflict between localized and itinerant approaches to
the classical 3d magnets\cite{Moriya_SCR,Gyorffy,Kubler_book}. 
The solution of this contradiction
was found in recognizing that the itinerant electrons can build well-formed
atomic moments participating in local transversal magnetic fluctuations.
On this basis it is possible to employ the DFT calculations for mapping an itinerant electron 
system with well-formed atomic moments onto the 
Heisenberg Hamiltonian \cite{Szilva2023,Mankovsky_parameters,Turek_parameters,Fawcett1988,Dai2012}. 

There is a straightforward DFT-based method to examine if
well-formed atomic moments are present in a given material.
It consists in carrying out self-consistent calculations
of various magnetic configurations. If such calculations converge to the magnetic states 
with close values of atomic moments the picture of well-defined atomic moments is justified.     
We will apply this method to UPS.

Another aspect of the dual nature of the 5f electron systems is the fact that 
the DFT methods often do not provide an accurate enough description of the
intraatomic 5f electron correlations.
Among the methods to improve on this is the DFT+$U$ method \cite{Anisimov1997}. 
We will employ both DFT and DFT+$U$ methods.

An important aspect of the physics of the U based materials is the multi-orbital 
nature
of the 5f electron system. A special form of the dual treatment of the U compounds was suggested \cite{Zwicknagl2003,Zwicknagl2016}
that consists 
in different treatment of the 5f electrons occupying different orbitals: Two electrons are 
treated as localized and not hybridizing with other electron states whereas the rest of the
5f electrons is treated as itinerant. We will refer to this form of duality as 
two-5f-electron-groups (T5FEG) duality.
This duality treatment was applied to UPS in Ref.~\onlinecite{Schulze2017}. 
This approach was successfully used in the study of
a series of UM$_2$Si$_2$ (M=Pd,Ni,Ru,Fe) compounds \cite{Amorese2020}. 
We will relate the results of our calculations to the assumptions of 
the T5FEG-duality approach. 

\section{Method of calculation}
\label{Sec_Method of calculations}

The calculations are performed with the augmented spherical waves (ASW)
method \cite{Williams1979,Eyert2012}
generalized to deal with 
spin-orbit coupling (SOC)
\cite{Sandratskii1998}.  The generalized gradient approximation (GGA) to the
energy functional \cite{Perdew1996} is employed in the calculations.
The DFT+$U$ method in the form suggested by Dudarev {\it et al.} \cite{Dudarev1998}
was applied to examine the influence of the on-site correlation of the U 5f electrons
on the magnetic moments and energies of the magnetic configurations.
The most of calculations
were performed with {\bf k}-mesh 20$\times$20$\times$20.
\textcolor{black}{
This allowed to reach the convergence of the energy differences between magnetic
configurations of 0.01~mRy per U atom.
}

An important quantity of the DFT+$U$
approach is the orbital density matrix $n$ of the correlated atomic states.
It enters the method
with the prefactor $U_H$ leading to the orbital dependence of the electron potential \cite{Dudarev1998}
\textcolor{black}{
\begin{equation}
V_{m,m'}^{s,s'}=-U_H(n_{m,m'}^{s,s'}-\frac{1}{2}\delta_{m,m'}\delta_{s,s'}).
\label{Eq_Vmm}
\end{equation}}
In the paper we work in the basis of complex spherical harmonics $Y_{lm}$.
The orbital dependence of the potential is given by the dependence on the magnetic quantum number $m$.
\textcolor{black}{
The diagonal elements $n_{m,m}^{s,s}$ of the orbital density matrix give the occupations of the orbitals
corresponding to quantum numbers
$m$ and spin projections $s$. 
}

We calculate the vectors of spin $\mathbf{m}_s^\nu$ and orbital $\mathbf{m}_{o}^\nu$ moments
of the $\nu$th atom as
\begin{equation}
\mathbf{m}_s^\nu=\sum_{\mathbf{k}n}^{occ}\int_{\Omega_{\nu}}\psi_{\mathbf{k}n}^\dag\mathbf{\sigma}
\psi_{\mathbf{k}n}d\mathbf{r}
\end{equation}
\begin{equation}
\mathbf{m}_{o}^\nu=\sum_{\mathbf{k}n}^{occ}\int_{\Omega_{\nu}}\psi_{\mathbf{k}n}^\dag\mathbf{\hat{l}}
\psi_{\mathbf{k}n}d\mathbf{r}
\end{equation}
where $\mathbf{\sigma}=(\sigma_x,\sigma_y,\sigma_z)$ is the vector of Pauli matrices
and $\mathbf{\hat{l}}=(\hat{l}_x,\hat{l}_y,\hat{l}_z)$ is the operator of orbital angular momentum, 
$\psi_{\mathbf{k}n}$ is
the wave function of the Kohn-Sham state corresponding to wave vector $\mathbf{k}$
and band index $n$. The sum is taken over occupied states.
The integrals are carried out over $\nu$th atomic sphere.

Due to the orbital-dependent potential term [Eq.~(\ref{Eq_Vmm})]
the occupied orbitals tend to lower their energies
whereas empty orbitals tend to increase their energies.
This feature makes the DFT+$U$ approach an adequate tool for the study of
the enhancement of the orbital magnetic moment \cite{Solovyev1998}.
We will study the dependence of the selected fundamental properties on $U_H$
aiming at obtaining the model describing the experimental results. 

The elements of the $n$ matrix determine the value of the OM \cite{Sandratskii2021}.
The $z$ component of the OM, $m_{oz}$, is determined by the diagonal elements
\textcolor{black}{
\begin{equation}
m_{oz}=\sum_s\sum_{m=-3}^3 m \: n_{m,m}^{s,s}=\sum_s\sum_{m=1,2,3} m\:(n_{m,m}^{s,s}-n_{-m,-m}^{s,s}).
\label{Eq_m_oz}
\end{equation}} 

We will perform calculations for four magnetic configurations: two AFM and two FM structures with 
moments parallel to the $z$ and $x$ axes. The configurations will be labeled as AFM$_Z$, AFM$_X$,
FM$_Z$, and FM$_X$.

In the projection of the quantities on an axis we will always chose the direction of the axis parallel to
the direction of the total moment. Therefore the projection of the orbital moment will be positive, and
the projection of the spin moment will be negative. Accordingly, the majority spin occupation corresponds to
the electron states with negative spin projection.

\section{Results and discussion}
\label{sec_Results and discussion}

\subsection{The results of the GGA calculations} 
\begin{table}[h]
\caption{Energies and U magnetic moments for four magnetic configurations
calculated with the GGA potential.
$m_s$, $m_o$ and $m_t$ are the values of spin, orbital and total moments
in units of $\mu_B$. The directions of the spin and orbital moments are opposite to
each other. The direction of the total moment is parallel to the direction of the 
orbital moment.}
\begin{center}
\begin{tabular}{c c c c c}
\hline
\hline
\hspace{5cm}& E (mRy/U) & $\;\;\;m_s\;\;\;$  & $\;\;\;m_o\;\;\;$  & $\;\;\;m_t\;\;\;$ \\
\hline
AFM$_Z$ &0    & 1.92  &  2.77& 0.85\\
AFM$_X$ &0.54 & 1.93  &  2.68& 0.75\\
FM$_Z$  &0.59 & 1.78  &  2.75& 0.97\\
FM$_X$  &1.27 & 1.79  &  2.62& 0.83\\
\hline
\label{table1}
\end{tabular}
\end{center}
\end{table}
We begin with the discussion of the results of the GGA calculations. 
The values of the energies and U magnetic moments 
are collected in Table~\ref{table1}.
In agreement with experiment, 
the ground state magnetic configuration is AFM$_Z$. Therefore, both the AFM magnetic structure and the 
easy $c$ axis are captured correctly. The total U moment of 0.85~$\mu_B$ is, as expected, too small.

\begin{figure}[t]
\includegraphics*[width=4cm]{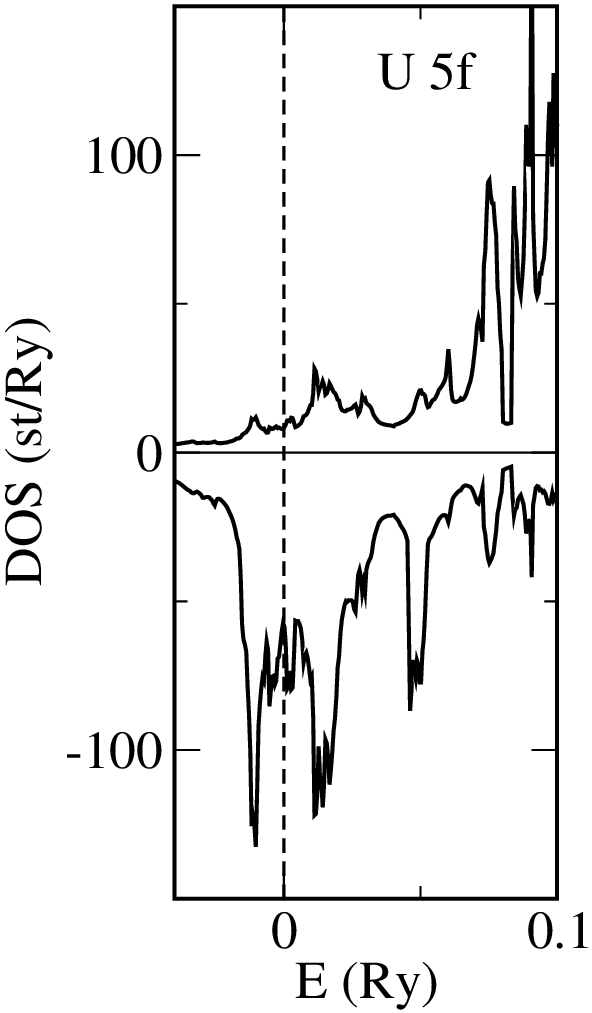}
\caption{Spin projected U 5f DOS for the AFM$_Z$ configuration. GGA calculation. 
}
\label{fig_DOS_AFMZ}
\end{figure}
\begin{figure}[t]
\includegraphics*[width=8cm]{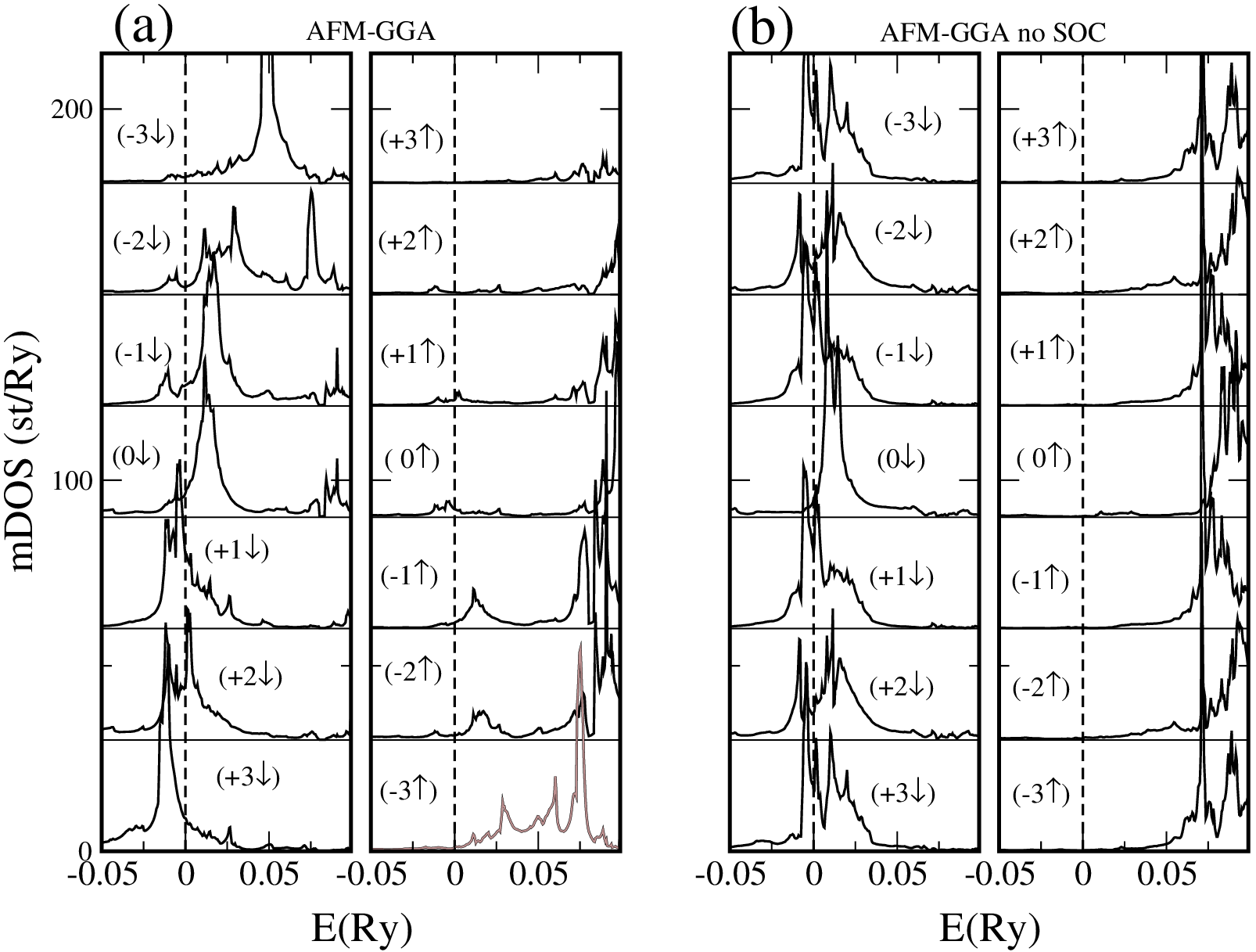}
\caption{$m$ and spin projected U 5f DOSs for the AFM$_Z$ configuration.
(a) The GGA calculation with SOC taken into account. (b) The GGA calculation without SOC.
}
\label{fig_mDOS_GGA}
\end{figure}
In Fig.~\ref{fig_DOS_AFMZ} we show the spin projected U 5f DOSs for the AFM$_Z$ configuration.
The origin of the spin moment is well seen: 
The spin splitting of the
states and, as a consequence, different occupation of the spin-up and spin-down 5f orbitals determines the value 
of the spin moment. However, the spin-projected DOSs provide no information on the origin and value of the 
OM.
To visualize the formation of the OM
we need $m$-resolved partial DOSs. They are presented in Fig.~\ref{fig_mDOS_GGA}(a). 
We see that the occupation of the 5f orbitals corresponding to positive $m$ values 
in the spin-down channel is much higher than
the occupation of the orbitals corresponding to negative $m$ values. This $\pm m$ polarization of the
mDOSs is the origin of the orbital magnetic moments [see Eq.~(\ref{Eq_m_oz})]. 
To understand deeper the interplay between spin and orbital components of the atomic moment
we performed the calculation of the AFM structure neglecting SOC.
We obtained spin moment of 2.16~$\mu_B$ and zero OM. 
Considering 
SOC-free mDOSs (Fig.~\ref{fig_mDOS_GGA}(b)) we see 
strong spin 
polarization of the U 5f orbitals whereas the $\pm m$ polarization 
is absent. (A detailed discussion of the symmetry properties of the $m$ projected DOSs
and occupation matrices $n$ can be found in Ref.~\onlinecite{Sandratskii2023}).

Continuing the analysis of the GGA results (Table~\ref{table1}) 
we notice that rather large atomic spin and orbital
moments are obtained for all four magnetic configurations. The values of the corresponding moments in 
all configurations are relatively close to each other.
This gives us the basis for the conclusion that the U magnetic moments in UPS can be characterized as 
rather well formed. On the other hand, the variation of the values of both spin and 
orbital moments between the configurations is not negligible and reflects
the influence of interatomic interactions, in particular interatomic  hybridization.

Concluding this section we remark that in the GGA calculations Bruno's relation
connecting MAE and OMA is fulfilled: the easy axis corresponds 
to the direction with a larger OM. 
In Sec.~\ref{sec_Bruno_rule} we will see that in the GGA+$U$ calculations this relation is not 
obtained
for some $U_H$ values. 

\subsection{Results of the GGA+$U$ calculations}
\subsubsection{Atomic moments as functions of $U_H$}
\label{sec_m_of_U}
Since the GGA calculations underestimate substantially the value of the atomic moment,
our next step is to carry out the GGA+$U$ calculations aiming at improved agreement
with experiment in this respect. 
Of course the GGA+$U$ calculations change also
the energies of the magnetic configurations. Therefore the question
whether the AFM$_Z$ configuration remains the lowest in energy 
is crucial for the realization of our goal to obtain the
model describing the selected set of the properties.

\begin{figure}[t]
\includegraphics*[width=8cm]{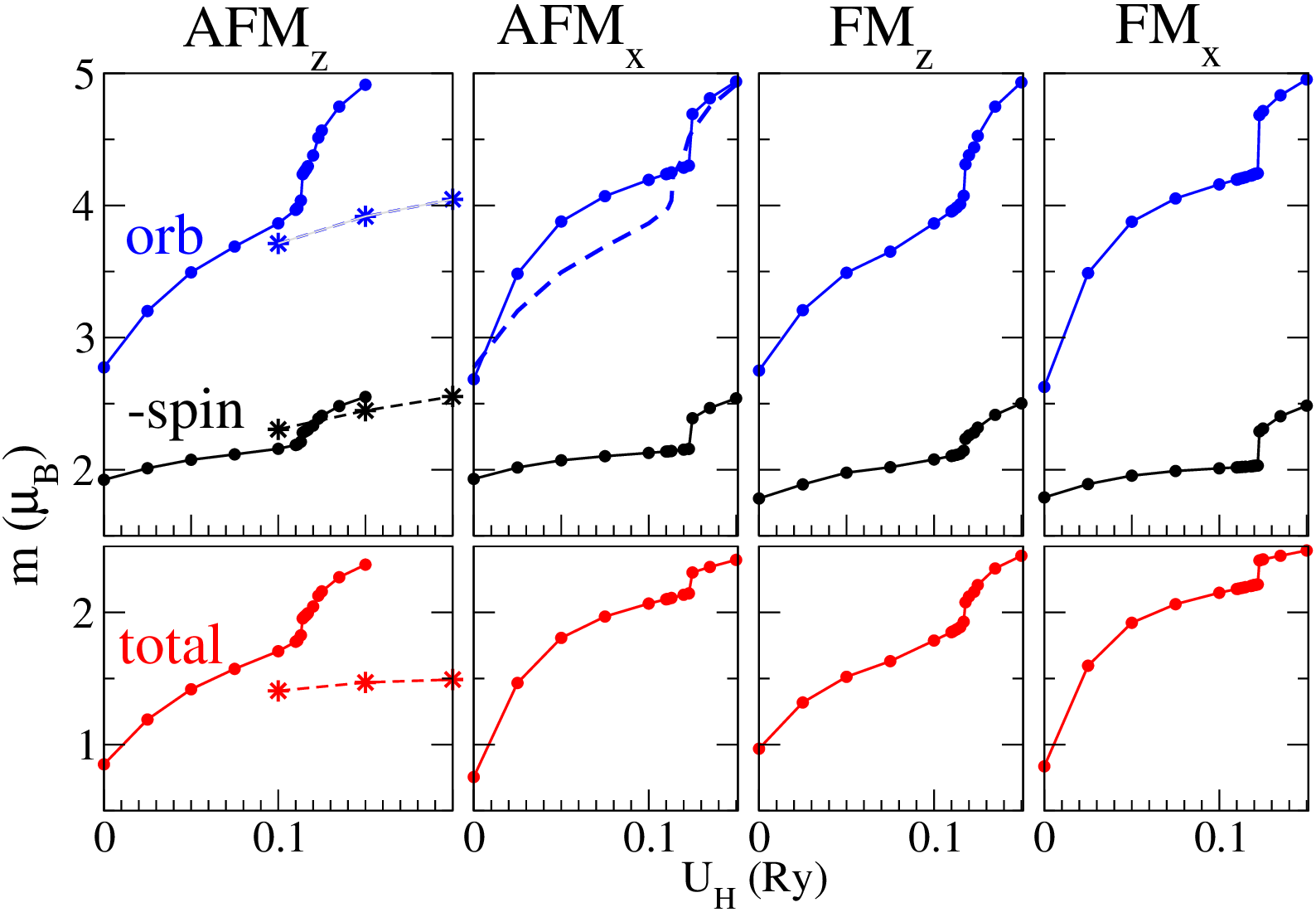}
\caption{The $U_H$ dependence of the spin, orbital and total U magnetic moments for four 
magnetic configurations. The broken line in the AFM$_X$ subgraph is the copy of the 
$m_o(U_H)$ dependence for the AFM$_Z$ configuration. 
\textcolor{black}{
The asterisks in the AFM$_Z$ subgraph show the data obtained in the T5FEG-model simulation
(Sec.~\ref{sec_T5FEG}).
}
}
\label{fig_m_of_U}
\end{figure}

In Fig.~\ref{fig_m_of_U} we present the $U_H$ dependence of the U moments for four magnetic configurations. 
Both spin and orbital moments for all 
configurations increase monotonously
with increasing $U_H$.
The OM grows faster than the spin moment leading to 
the monotonous increase of the total moment and improving agreement with experiment. 
For the AFM$_Z$ configuration and $U_H$$>$0.12~Ry the value of the total U moment 
exceeds 2~$\mu_B$.

\begin{figure}[t]
\includegraphics*[width=6cm]{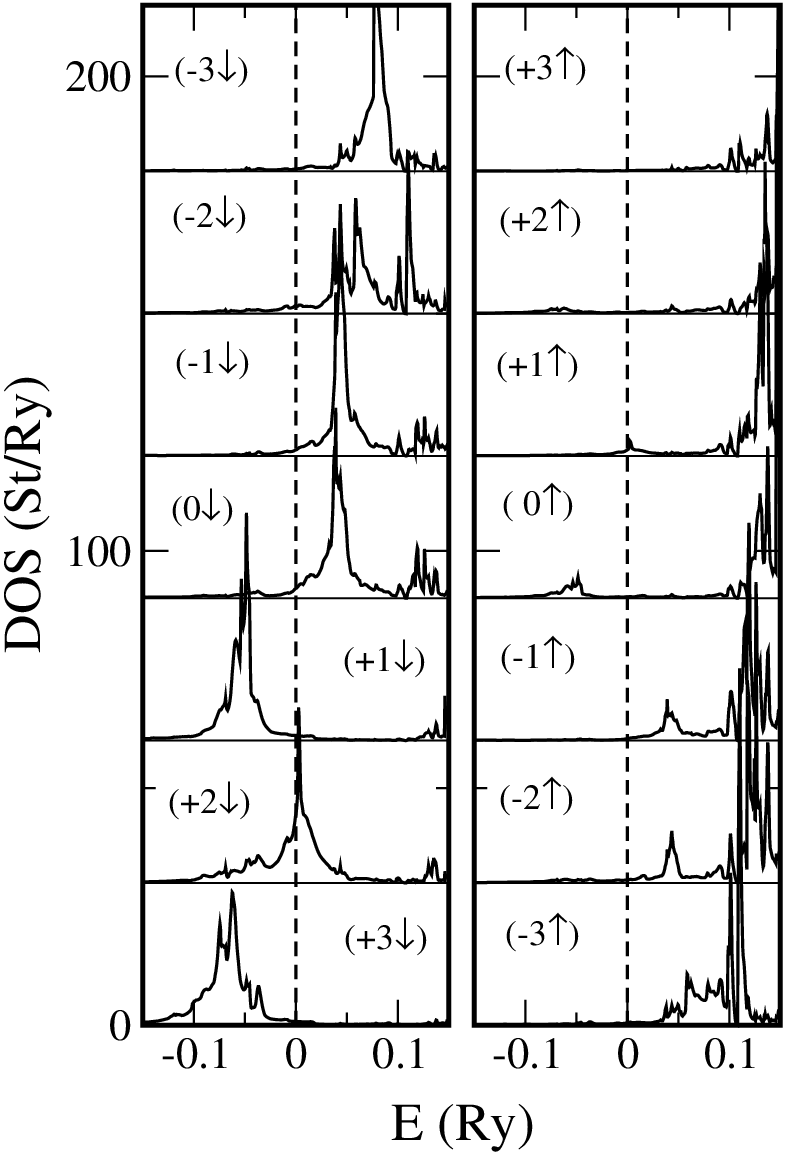}
\caption{The $m$ and spin projected U 5f DOSs for the AFM$_Z$ configuration
calculated with GGA+$U$ method and $U_H$=0.1~Ry.
}
\label{fig_mDOS_U1}
\end{figure}

In Fig.~\ref{fig_mDOS_U1}, we show the mDOSs of the AFM$_Z$ configuration calculated with 
$U_H$=0.1~Ry. The comparison with the result of the GGA calculation shows that the energies of the 
$m=3$ and $m=1$ orbitals lie now distinctly below the Fermi level ($E_F$) and corresponding partial mDOSs are filled.
The enhanced $\pm m$ polarization leads to the increase of the U OMs.
To explain the difference in the $U_H$ dependences of the spin and orbital moments 
we compare partial U 5f mDOSs obtained with GGA (Fig.~\ref{fig_mDOS_GGA}) 
and GGA+$U$ (Fig.~\ref{fig_mDOS_U1}) calculations. 
As seen in Fig.~\ref{fig_mDOS_GGA}, 
already for $U_H$=0, the spin polarization of the 5f orbitals is large while
their $\pm m$-polarization is less manifested [Fig.~\ref{fig_mDOS_GGA}(a)]. 
Although $U_H$ tends to increase both 
polarizations, the 
large initial spin polarization limits its further increase.

Another feature obtained for all four magnetic configurations is the 
discontinuities in the $m(U_H)$ dependences at $U_H$ values close to 0.12~Ry.
The position of the singularity changes somewhat from case to case. 
\begin{figure}[t]
\includegraphics*[width=4cm]{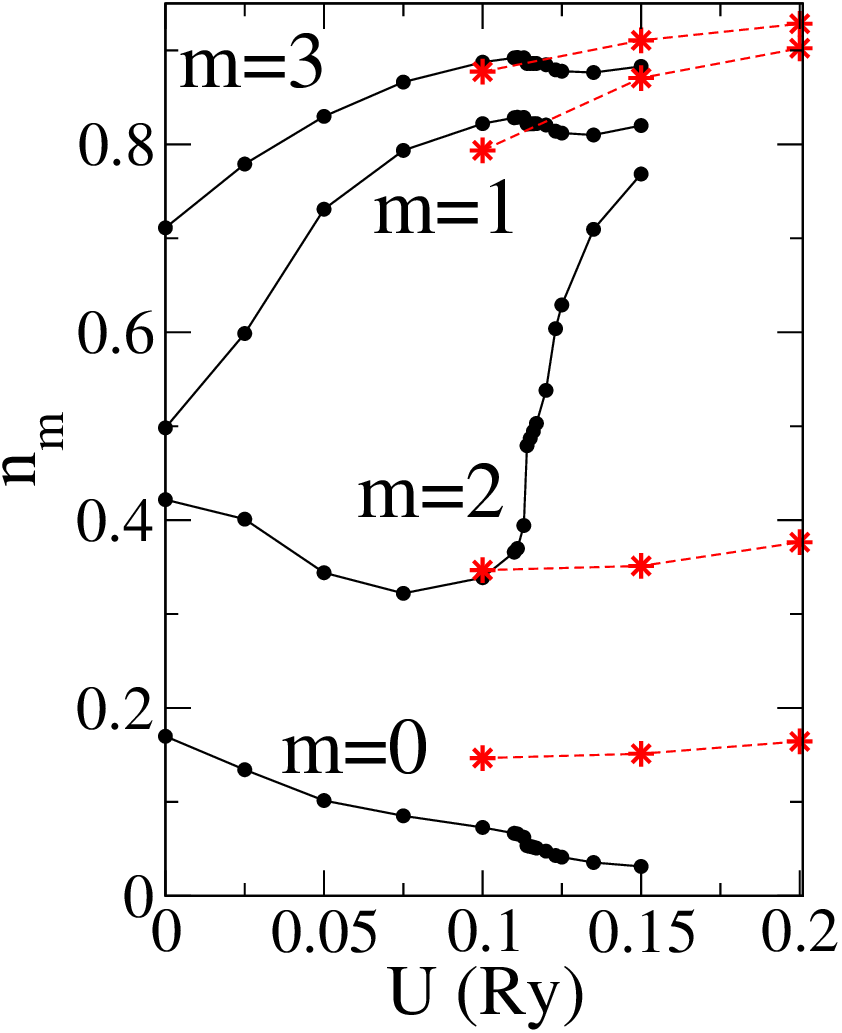}
\caption{The occupation numbers $n_m$$\equiv$$n_{m,m}$ for $m=0,1,2,3$
and spin-majority channel for the AFM$_Z$ configuration.  
The data presented by asterisks will be discussed in relation 
with the simulation of the T5FEG model (Sec.~\ref{sec_T5FEG}).
}
\label{FIG_occ__of_U}
\end{figure}
\textcolor{black}{
To understand the nature of the discontinuous behavior 
we analyze in the case of the AFM$_Z$ configuration the $U_H$ dependence 
of the occupation numbers $n_{m,m}$ for $m=0,1,2,3$
and the spin-majority channel (Fig.~\ref{FIG_occ__of_U})\cite{comment_electron_number}.  
Obviously, the origin of the discontinuity is in the properties of the 
occupation number for $m=2$. This is the U~5f orbital heavily present at the Fermi
level. We remark that strong increase of the $m=2$ occupation is accompanied by some
decrease of other occupation numbers. The sum of the occupation numbers for all $m$
and both spin projections
changes from 2.52 for $U_H=0.15$~Ry to 2.78 for $U_H=0.1$~Ry.  
}

In general, the shapes of the $U_H$ dependences for corresponding
moment types
are rather similar in all four cases.
This reveals that the formation
of the moments is predominantly a local atomic effect while
the change of the relative orientation of the atomic moments and their 
orientation with respect to the lattice have a smaller influence.
Most importantly, 
the conclusion that UPS 
is the material with well-formed U atomic moments 
remains valid also in the GGA+$U$ calculations for all $U_H$ values.

\begin{figure}[t]
\includegraphics*[width=8cm]{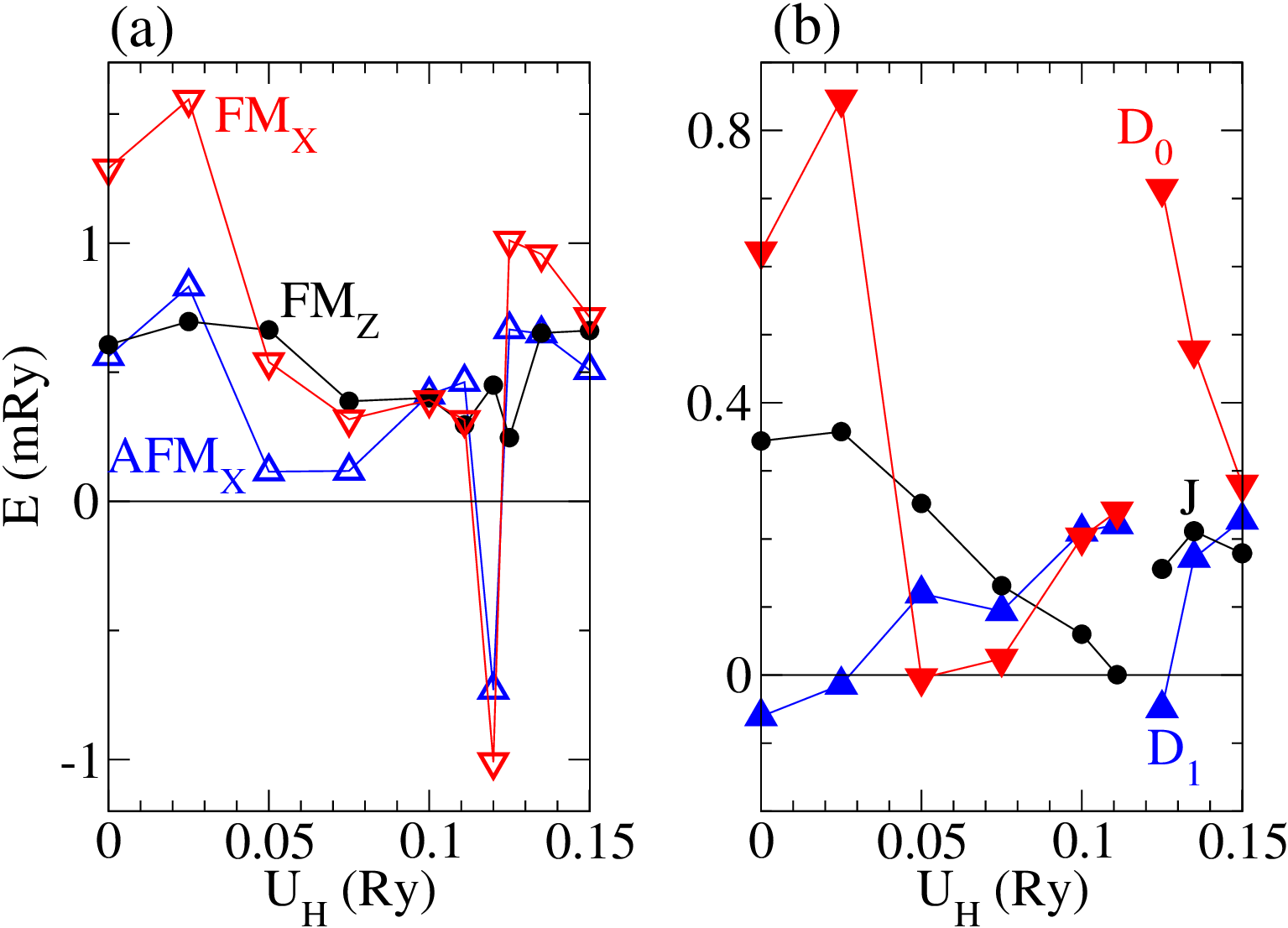}
\caption{(a) Energies of the AFM$_X$, FM$_Z$ and FM$_X$ configurations 
as functions of $U_H$ counted
from the energy of the AFM$_Z$ configuration.
(b) Parameters of the model Hamiltonian of interacting atomic moments
as functions of $U_H$:
isotropic exchange $J$, single-site anisotropy $D_0$, two-site anisotropy
(anisotropic exchange) $D_1$.
}
\label{fig_E_of_U}
\end{figure}

\subsubsection{Energies of magnetic configurations as functions of $U_H$}
\label{sec_E_of_U}

In Fig.~\ref{fig_E_of_U}(a) we show the $U_H$ dependence of 
the energies of the AFM$_X$, FM$_Z$ and FM$_X$ configurations counted 
from the energy of the AFM$_Z$ configuration. 
We see that for all but one values of the $U_H$ mesh the AFM$_Z$ configuration is
the lowest in energy. This means that the agreement with experiment concerning both the 
magnetic structure and the magnetic easy axis obtained for $U_H$=0 is a robust result 
preserved for most of the $U_H$ values. 
At the same time, as we have seen in Sec.~\ref{sec_m_of_U}, the use of the GGA+$U$ method improves the agreement 
with experiment concerning the value of the magnetic moment. 

We suggest the following explanation 
of the deviating result for $U_H$=0.12~Ry.
This $U_H$ value
lies in the region of the discontinuities in the $U_H$ dependences of magnetic moments (Fig.~\ref{fig_m_of_U}). 
Importantly, it is above the discontinuity point for AFM$_Z$ and below it for AFM$_X$ and FM$_X$. 
Apparently, the comparison of the energies of two
states, one of which is before and the other after the discontinuous transformation, 
can lead to the disareement with experiment.
The assumption that different 
states of the system are characterized by the same
$U_H$ value has been widely and successfully used. However, it does not have the status of a mathematically 
proven statement. 
Since the screening of the
Coulomb interaction can change with the change of magnetic configuration,
somewhat different 
$U_H$ values 
should be expected for different configurations.
In the region of the discontinuities
this effect is expected to be enhanced.

We can summarize this issue as follows.
Our analysis is focused on the trends in the $U_H$ dependences.
The comparison of the energies of different states of the system obtained with the same $U_H$
value is a common practice proved to be reliable in many physical problems. In our case, we obtained
agreement with experiment concerning both the magnetic structure and magnetic anisotropy
for most of the $U_H$ values. For the $U_H$ values above the discontinuity points we have 
good agreement with experiment also in the value of the atomic moment
On the other hand, our results show that in the regions of 
discontinuous behavior
an extra caution is required.

Continuing the analysis of the energy dependences $E(U_H$) 
we notice that
there are intersections of the functions corresponding to different 
magnetic configurations [Fig.~\ref{fig_E_of_U}(a)]. 
Since the energies of the magnetic configurations are determined by both exchange 
interaction and magnetic anisotropy, these results show that relative strength of 
the interactions varies with the variation of $U_H$. 
In particular, such a behavior shows that FM and AFM structures for some $U_H$ values 
have different easy axes. This reveals the importance
of the two-site anisotropy (anisotropic exchange). 
We will return to this issue in Sec.~\ref{sec_calc_parameters} where the estimates of the 
anisotropy and exchange parameters are reported.

\subsection{Interpretation of the in-field experiment}
\subsubsection{Experimental facts and general considerations}
The high field magnetization measurements \cite{Schulze2012} show
that for the maximal applied magnetic field of 50 T 
the induced moments for $B||z$ and 
$B||x$ have comparable values of, respectively, 
~1.5 $\mu_B$ and ~1.1 $\mu_B$ per U atom. 
These values, although large, are distinctly
smaller than the ground-state atomic moments.
The shapes of the $B||x$ and $B||z$ magnetization
curves are very different. For the description of all details of the 
experimental curves we refer the reader
to the original paper \cite{Schulze2012}  and review \cite{Mydosh2017}. 
We will not attempt to theoretically reproduce 
the experimental dependences in their full
complexity. The main difference between the two experimental curves which we aim to understand is
that for the $B||z$ field there is a strong increase of the magnetization by about 1~$\mu_B$ 
in a relatively narrow field interval around 30~T 
revealing the presence of one or even two MMTs whereas
in the $B||x$ case such a strong feature is absent.

This combination of properties reflects a certain relation between the interatomic exchange
energy, magnetic anisotropy energy and in-field Zeeman energy. We will
estimate corresponding energy scales and relate our estimations to the experimental situation. 
Since the calculations 
have shown the presence of well-formed atomic moments (Sec.~\ref{sec_m_of_U}) our interpretation
of the high field experiment
considers the reorientation of the atomic moments
as predominant factor. 

It is worthwhile to briefly discuss the limiting cases where one of the three competing energy contributions 
is distinctly dominating. 
(i) If the AFM exchange interaction is dominant, a considerable distortion of the AFM structure
cannot take place and a large induced magnetization is not expected 
for any field direction.
(ii) If the MAE is dominant, the large induced moment for $B||x$ is not possible. For field $B||z$
exceeding the strength of the exchange interaction, the discontinuous spin-flip transition to the 
field-induced ferromagnetic state parallel to the $z$ axis is expected. 
In this spin-flip state the momnent per U atom 
would exceed 2 $\mu_B$ that is larger than the observed value of the induced moment. 
(iii) If the Zeeman energy dominates, 
the spin-flip transition to the field-induced ferromagnetism
is expected for any field direction. 

All these limiting scenarios do not agree with experimental observations 
where large induced moments are 
obtained for both field directions 
and the induced moments are smaller than atomic moments. 
Therefore, the conclusion is that we deal with an interplay of different 
interactions having comparable energy scales \cite{Delgado2023}.
In Sec.~\ref{sec_calc_parameters} we report
an estimation of the scales of the 
exchange energy, MAE, and Zeeman energy (see Secs.~\ref{sec_calc_parameters}, \ref{sec_Zeeman}). 

\subsubsection{Calculation of exchange interaction and magnetic anisotropy parameters}
\label{sec_calc_parameters}

The available energies of four magnetic configurations give three energy differences and 
allow us to perform an estimation of 
three interaction parameters of the model Hamiltonian of interacting atomic moments
\begin{equation}
H=\sum_{ij} \hat{\bm S}_iA^{(i,j)} \hat{\bm S}_j^T
\end{equation}
where $A^{(i,j)}$ are 3$\times$3 interaction matrices,
$T$ means matrix transposition,
$\hat{\bm S}_i$ is the unit vector in the direction of the $i$th atomic moment.
To do the parameter estimation efficiently
it is important to perform the symmetry analysis of the
$A^{(i,j)}$ matrices. If the SOC is not taken into account, 
the matrices have the scalar form 
$J\left(\begin{array}{ccc}
1 & 0 & 0 \\
0 & 1 & 0 \\
0 & 0 & 1  
\end{array}\right)$ 
which reflects the isotropic character of interatomic 
exchange interactions and the absence of magnetic anisotropy. 
In the relativistic
case, only the hermicity of the matrices is guaranteed by the general principles. 
The lattice symmetry imposes further restrictions on the form of the matrices.
If $\{\alpha|\bm{\tau}_\alpha\}$ is the symmetry operation of the lattice 
consisting from rotation $\alpha$ and translation $\bm{\tau}_\alpha$,
it imposes the following constraint on the 
interaction matrices 
\begin{equation}
A^{(i,j)}=\alpha^TA^{(i_\alpha, j_{\alpha, \bm{R}})}\alpha
\label{eq_sym_matr}
\end{equation}
where the U sublattices $i_\alpha$, $j_\alpha$ and lattice vector $\bm{R}$
are defined by the action of operation  $\{\alpha|\bm{\tau}_\alpha\}$ on 
positions of atoms $i$ and $j$ (see Ref.~\onlinecite{Sandratskii2016} for 
the detailed description).

The crystal lattice of UPS is characterized by 16 point operations.
Applying them according to Eq.~(\ref{eq_sym_matr}),  
for the single-site matrix we obtain the simple form
\begin{equation}
A^{(i,i)}=\left(\begin{array}{ccc}
0 & 0 & 0 \\
0 & 0 & 0 \\
0 & 0 & D_0 
\end{array}\right) 
\end{equation}
where $D_0$ is the single-site anisotropy parameter.
For the interaction matrices between atoms 0 and 1 of the two U layers
(Fig.~\ref{Fig_lattice})
we get 
\begin{equation}
A^{(0,1)}=\left(\begin{array}{ccc}
j_1-\frac{1}{3}d_1 & b & c \\
b & j_1-\frac{1}{3}d_1 & c \\
c & c &j_1+\frac{2}{3}d_1  
\end{array}\right) 
\end{equation} 
where $j_1$ is isotropic exchange parameter, $d_1$ is two-site anisotropy parameter or,
equivalently, anisotropic-exchange parameter, $b$ and $c$ are
real numbers.
The interaction matrices between atom 0 and atoms 1-4 
are transformed
to each other by symmetry operations and therefore are determined by the same set of
parameters. 
For atomic pairs (0,2), (0,3) and (0,4) we get, respectively:
\begin{widetext}
\begin{equation}
\left(\begin{array}{ccc}
j_1-\frac{1}{3}d_1 & -b & c \\
-b & j_1-\frac{1}{3}d_1 & -c \\
c & -c &j_1+\frac{2}{3}d_1  
\end{array}\right), 
\left(\begin{array}{ccc}
j_1-\frac{1}{3}d_1 & b & -c \\
b & j_1-\frac{1}{3}d_1 & -c \\
-c & -c &j_1+\frac{2}{3}d_1  
\end{array}\right), 
\left(\begin{array}{ccc}
j_1-\frac{1}{3}d_1 & -b & -c \\
-b & j_1-\frac{1}{3}d_1 & c \\
-c & c &j_1+\frac{2}{3}d_1
\end{array}\right).
\end{equation}
\end{widetext}

The same symmetry properties are valid for the interaction matrices between atom 0
and atoms 5-8 of the lower U layer (Fig.~\ref{Fig_lattice}). While the sets 
of parameters for the two 
groups of interaction matrices are not equivalent by symmetry we introduce for atoms 5-8 
the notations
$j_2$ and $d_2$ instead of $j_1$ and $d_1$. 

The energies of the four magnetic configurations in terms of the parameters of the 
interaction matrices take the form
\begin{equation}
\begin{split}
&E(AFM_Z)=D_0-J-\frac{2}{3}D_1\\
&E(AFM_X)=-J+\frac{1}{3}D_1\\
&E(FM_Z)=D_0+J+\frac{2}{3}D_1\\
&E(FM_X)=J-\frac{1}{3}D_1
\label{eq_4_energies}
\end{split}
\end{equation}
where $J=4(j_1+j_2)$ and $D_1=4(d_1+d_2)$.

Equations~(\ref{eq_4_energies})  allow us to uniquely determine parameters $J$, $D_0$, $D_1$.
They are presented as functions of $U_H$ in Fig.~\ref{fig_E_of_U}(b).
The anomalous values at $U_H$=0.12~Ry 
resulting from the anomalous properties of the energy differences discussed above
are not presented.
In general, the parameters have comparable scales. 
Because the dependences of the parameters on $U_H$
are different their relative strength also varies.
For example, for $U_H$=0 and $U_H$=0.025~Ry, the single-site anisotropy
dominates distinctly. 
At $U_H$=0.1~Ry, both anisotropy
parameters practically coincide. 
For the upper part of the $U_H$ values
lying above the 
discontinuity region the values of all three interaction parameters are similar. This
region is of the main interest for us since here the model reproduces correctly all three
fundamental ground state properties: magnetic structure, magnetic easy axis, the value of the 
magnetic moment. 
The $U_H$-dependent relation between the values of the single-site and two-site anisotropies 
reflects the importance of both single-site and inter-site processes and can be treated 
as one of the manifestations
of the localization-delocalization duality.

\subsubsection{Zeeman energy and interpretation of experiment}
\label{sec_Zeeman}
Let us make an estimation of the relevant Zeeman energy. If we take magnetic field of 30~T
and the U magnetic moment of 2.5~$\mu_B$ we obtain the energy of $\sim$0.3~mRy 
per U atom
that is of the same scale as the three other estimated parameters [Fig.~\ref{fig_E_of_U}(b)].   

\begin{figure}[t]
\includegraphics*[width=5cm]{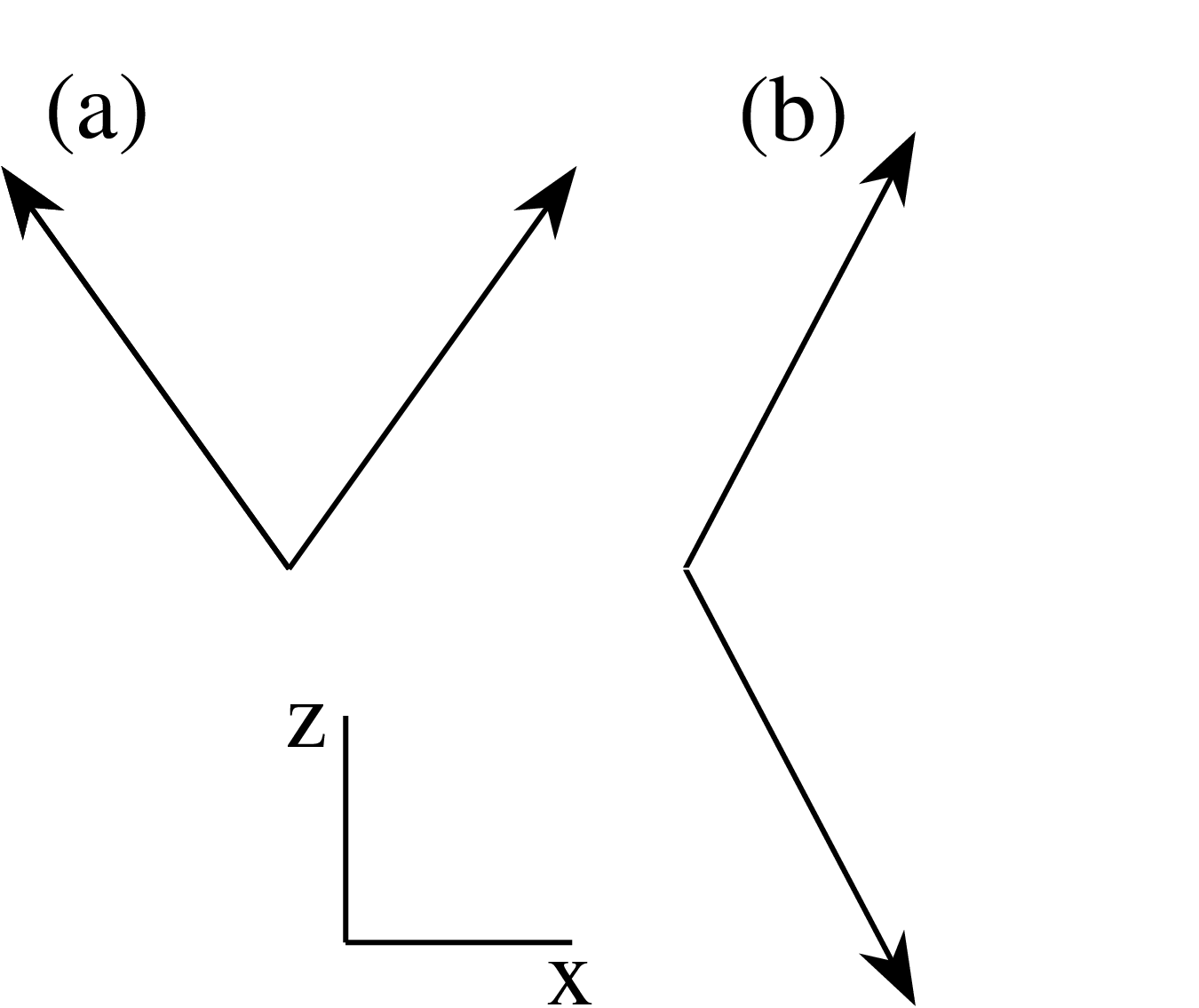}
\caption{Schematic presentation of canted magnetic structures in magnetic 
field parallel to the $z$ axis (a) and $x$ axis (b). In the ground state the 
the magnetic moments are collinear to the $z$ axis
and antiparallel to each other.
}
\label{fig_canting}
\end{figure}
The comparable values of different energy factors explain 
why large values of the moments are induced 
for both easy and hard field directions.
The induced magnetic configurations 
are canted magnetic structures with 
average magnetization
parallel to the direction of the field. A schematic presentation of two canted states
is given in Fig.~\ref{fig_canting}.
The canting means the deviation of the atomic moments from both the ground state AFM structure and 
the easy $c$ axis. The corresponding energy increase is compensated by the Zeeman
energy of magnetic moments. 

The close-to-discontinuity field dependence, i.e. MMT or spin-flop transition, for $B||z$ is the 
consequence of the MA since the continuous 
transition of the U moments with negative $z$ projection
towards the positive direction of the $z$ axis needs passing the $xy$ plane where 
the MA energy is high
(more about MMT in an AFM can be found in, e.g., 
Refs.~\onlinecite{Bogdanov2007,MMT1,MMT2}). 
For the canting of the moments toward the $x$ axis,
such an energy barrier does not appear.

Our interpretation of the in-field experiment is primarily based on the local-moment picture. Of course the
presence of the 5f states at the Fermi level contributes to 
the values of the moments and of the energies of magnetic configurations. The longitudinal scenario
of the strong magnetization change seems, however, improbable 
because the energy cost of the large change of the atomic moment value
is much higher than 
the energy gain from the Zeeman interaction of these moments with 
external field. For instance, in the GGA calculation the energy difference 
between AFM$_Z$ and nonmagnetic state is as high as 7.25~mRy/U.

\subsection{Relation metween MAE and OMA}
\label{sec_Bruno_rule}

We notice that our results violate the relation between MAE and OMA suggested by Bruno\cite{Bruno1989}:
the easy magnetic direction corresponds to the largest
value of the OM.
In our calculations, the OM of the AFM$_X$ configuration is for many $U_H$ values 
larger than the OM of the AFM$_Z$ configuration (Fig.~\ref{fig_m_of_U}) 
whereas the energy of the AFM$_Z$ configuration is the lowest (Fig.~\ref{fig_E_of_U}).
Our next step is to understand the origin of this result in the case of UPS.

The OM appears as a result of the disbalance in the occupation of the 
orbitals with opposite values of the magnetic quantum number, $\pm m$-polarization. 
If the SOC is not taken into account, the occupations of both orbitals are identical 
and both OM and MAE are zero.
Because of $\propto$$\,\sigma_Zl_Z$ term, the SOC tends to generate
the $m$-dependent shifts of the 5f orbital energies 
(see Ref.~\onlinecite{Sandratskii2017} for full expression of the SOC operator used 
in the calculations). 
For negative spin projection 
one expects the lowest energy position of the $m=3$ orbital and monotonously
increasing energy with decreasing $m$. 
The monotonous $m$-dependence of the energy positions leads 
to the monotonous variation
of the orbital occupations which directly influences the value of the orbital 
moment. This connection between energies of the orbitals and atomic orbital moment
is the physical basis of Bruno's rule.

\begin{figure}[t]
\includegraphics*[width=6cm]{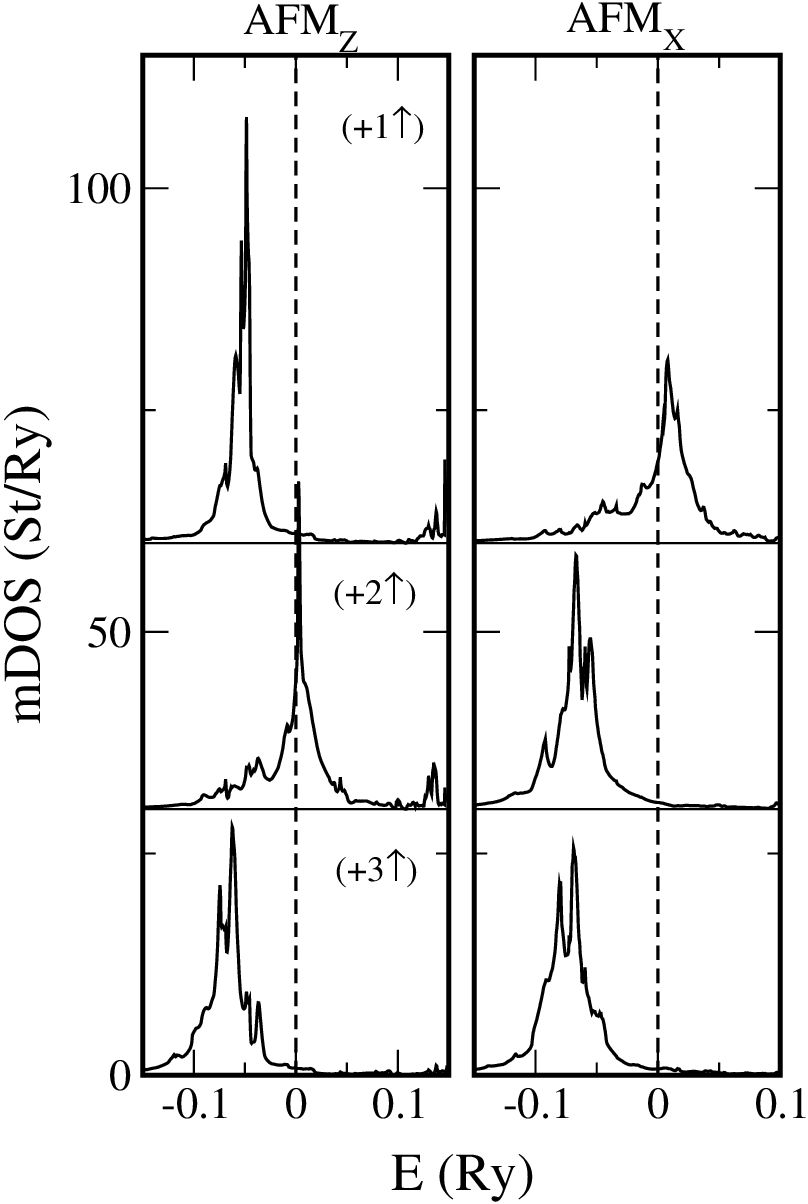}
\caption{Comparison of the low-energy mDOSs of the AFM$_Z$ and AFM$_X$ configurations
calculated with $U_H$=0.1~Ry.
}
\label{fig_mDOS_AFMzx_U1}
\end{figure}
However,
in the calculation for the ground state configuration AFM$_Z$ of UPS we obtained 
a nonmonotonous $m$-dependence of the energy positions of the $m$ projected DOSs
(see Fig.~\ref{fig_mDOS_AFMzx_U1}
for $U_H$=0.1~Ry). We connect this result 
with the hybridization of $m=3$ and $m=1$ orbitals that does not allow the intraatomic
SOC to split them. The filling of the $m=1$ orbital instead of the $m=2$ orbital reduces the value
of the OM. This disturbs the straightforward relation between energies of the orbitals and corresponding
OM.

Now let us turn to the AFM$_X$ configuration. The mDOSs and $m$-occupancies for the spherical 
harmonics defined with 
respect to the $z$ axis are not helpful for the analysis of the OMs parallel to the $x$ axis.
In Fig.~\ref{fig_mDOS_AFMzx_U1} we plot mDOSs of AFM$_X$ for the basis of the 5f functions defined with respect 
to the $x$ axis as quantization axis.
In this case we obtain a usual picture: the $m$-dependence of the energy positions and, 
therefore, $m$-occupancies are monotonous with respect to $m$.
We see here the competition between local tendency to the monotonous $m$-dependence and the 
influence of the electron delocalization results in desterbing simple relation between MAE and OMA.
This is not the first case where the Bruno's relation is not fulfilled\cite{Andersson2007}
On the other hand, 
this relation
is operative for numerous
materials, also for U compounds (see, e.g., Ref.~\onlinecite{Miyake2018}).
 
\subsection{Relation to the T5FEG-duality model}
\label{sec_T5FEG}

Our GGA+$U$ calculations give the results which can be treated as revealing the coexistence of 
localized and itinerant 5f states supporting the assumption of the T5FEG-duality model.
Indeed, in the ground state AFM$_Z$ configuration (Fig.~\ref{fig_mDOS_U1}) we have two 5f 
orbitals lying distinctly below $E_F$
while the states related to a third 5f orbital are in the Fermi energy region.
We emphasize that this result is obtained applying the same $U_H$ to all 5f orbitals.
In the spirit of the T5FEG model
it seems plausible to consider a different treatment of the two groups of the 5f orbitals
applying the GGA+$U$ term to only $m=3$ and $m=1$ orbitals\cite{comment_m01}.
In this
model calculation we were unable to reach the experimental value of the atomic moment. 
\textcolor{black}{
As seen in Fig.~\ref{fig_m_of_U}, in the $U_H$ interval from 0.1~Ry to 0.2~Ry both spin and orbital moments grow 
somewhat with increasing $U_H$.
However, these changes mostly compensate each other resulting in the total moment close
to 1.5~$\mu_B$ for the whole $U_H$ interval, that is distinctly smaller than the experimental value
and in strong contrats to the results of the GGA+$U$ calculations.
}

\textcolor{black}{
To understand this feature we compare the occupation numbers obtained in the T5FEG calculation with 
those obtained in the standard GGA+$U$ calculations (Fig.\ref{FIG_occ__of_U}). 
The crucial difference in the results of the 
two calculations is the absence in the T5FEG case of both the discontinuous behavior and fast increase of 
the occupation number of the $m=2$ obtained in the GGA+$U$ calculations. This increase 
is critical for getting the value
of the U moment close to the experimental one.
The sum of the occupation numbers for all $m$
and both spin projections
is 2.56 for $U_H=0.1$~Ry and 2.64 for $U_H=0.1$~Ry
what compares well with given above corresponding values from the GGA+$U$ calculations.
(For completeness, in Fig.~\ref{fig_mDOS_T5FEG} of Appendix~\ref{sec_mDOS_T5FEG}  
we compare mDOSs calculated with
GGA+$U$ and in the T5FEG simulation.)
}

This failure of the modified method is one more signature of the complexity of 
the multiple-orbital nature of 
the U 5f electron system in the U based materials. 
The neglect of the influence of the correlation on the $m=2$ orbital,
contributing to the formation of itinerant electron states,
results in the changes in the 5f electron system that does not allow to reach the
agreement with experiment.

\section{Conclusions}

The paper deals with the U based compound UPS. 
A wide variation of the 
properties of the U compounds demands the application of the material specific
DFT based methods to their theoretical study. In the case of UPS, the number of such 
studies is very scarce. In contrast, the experimental data are rich.
The goal of this paper is to contribute to filling this gap.
We start with identification of the DFT+$U$ based physical model
adequately describing the set of fundamental ground state properties: magnetic structure, magnetic
easy axis, the value of the U atomic moment. We demonstrate that the system possesses well formed
atomic moments able to participate in transversal magnetic fluctuations.

On this basis the high magnetic field experiment
is interpreted. This experiment shows the presence of a strong MMT for the field direction 
parallel to the easy axis in contrast to the hard direction where such a clear feature is absent.
On the other hand, comparable induced magnetization values are obtained for both easy and hard field directions. 
Within the framework of the suggested model, we interpret
this combination of properties as the result of the competition of three energy contributions:
exchange energy, MAE, and Zeeman energy. All three energies are estimated and shown
to have comparable values.

We notice and discuss the violation in the case of UPS of the Bruno's relation between MAE and OMA for
some of the $U_H$ values.

At all steps of the study we devote special attention to revealing
and emphasizing
the dual itinerant-localized  
properties of the material. 
Such an analysis is a useful tool for the in-depth study of the physics of the system. 
The obtained forms of
the duality are different: well defined atomic moments formed by the itinerant electrons,
interplay of the single-site and two-site anisotropies, strong localization of two of the 5f
electrons in contrast to the itinerant nature of the 5f electrons contributing to the states
around the Fermi level, intense Stoner continuum competing with spin wave formation. 

The paper contributes to the 
understanding that the 
wide range of complex behavior observed in U based materials
is the consequence of the multiple-orbital nature of the 5f electron system whose
properties are sensitive to numerous factors such as crystal structure,
ligand types, intraatomic electron correlation, strength and direction of the 
applied magnetic field. 

The continuation of the study of UPS in the following two directions 
appears to us of high interest.
\textcolor{black}{
The first is the numerical investigation of the magnetic excitations
by means of material-specific DFT-based calculation of transverse dynamic magnetic susceptibility 
(see, e.g., Refs.~\onlinecite{Buczek2011,Lounis2015,Skovhus2021})
aiming at exposing the reason for
the absence of the spin waves in the INS experiment\cite{Lee2018}. 
A possible reason for the absence of the spin waves is 
the presence of intensive low-energy Stoner continuum
leading to strong damping of the potential spin waves.
There are two factors telling us that 
such a continuum is expected in UPS. The first is 
the presence of the electron states corresponding to the U 5f orbitals in the Fermi 
level region [Fig.~\ref{fig_mDOS_T5FEG}(d)], both below and above $E_F$. 
Second, it is important that the magnetic structure is AFM and all electron
bands are double degenerate with pairs of states having equal energies and opposite $z$ projections
of the magnetic moments (see, e.g., Ref.~\onlinecite{Sandratskii2012} for a detailed discussion 
of the Stoner excitations in an AFM). 
Therefore, the numerous states participating in the low-energy
spin-flip transitions are available.
Another feature to remark is the presence of a narrow empty DOS peak close to $E_F$ 
[Fig.~\ref{fig_mDOS_T5FEG}(d)].
The electron transitions to this peak build the main part of the Stoner continuum.
It would be of high interest to compare theoretical TDMSs for various U compounds 
with available experimental INS data to obtain a general picture.}
We should, however,
remark that, in the case of the U based materials, the DFT-based calculation of the TDMA
is a very complex computational task since
it must include a consequent account for both strong SOC and local electron correlation governed 
by Hubbard parameter $U_H$. 

The second direction is the application of the methods 
of more advanced account
for electron correlation than DFT+$U$, in particular of the DFT plus dynamic mean field theory 
(DFT+DMFT) method\cite{Kotliar2006}. This problem is also numerically very challenging and 
computer resources demanding.

\appendix
\section{DOS in a wide energy interval}
\begin{figure}[t]
\includegraphics*[width=8cm]{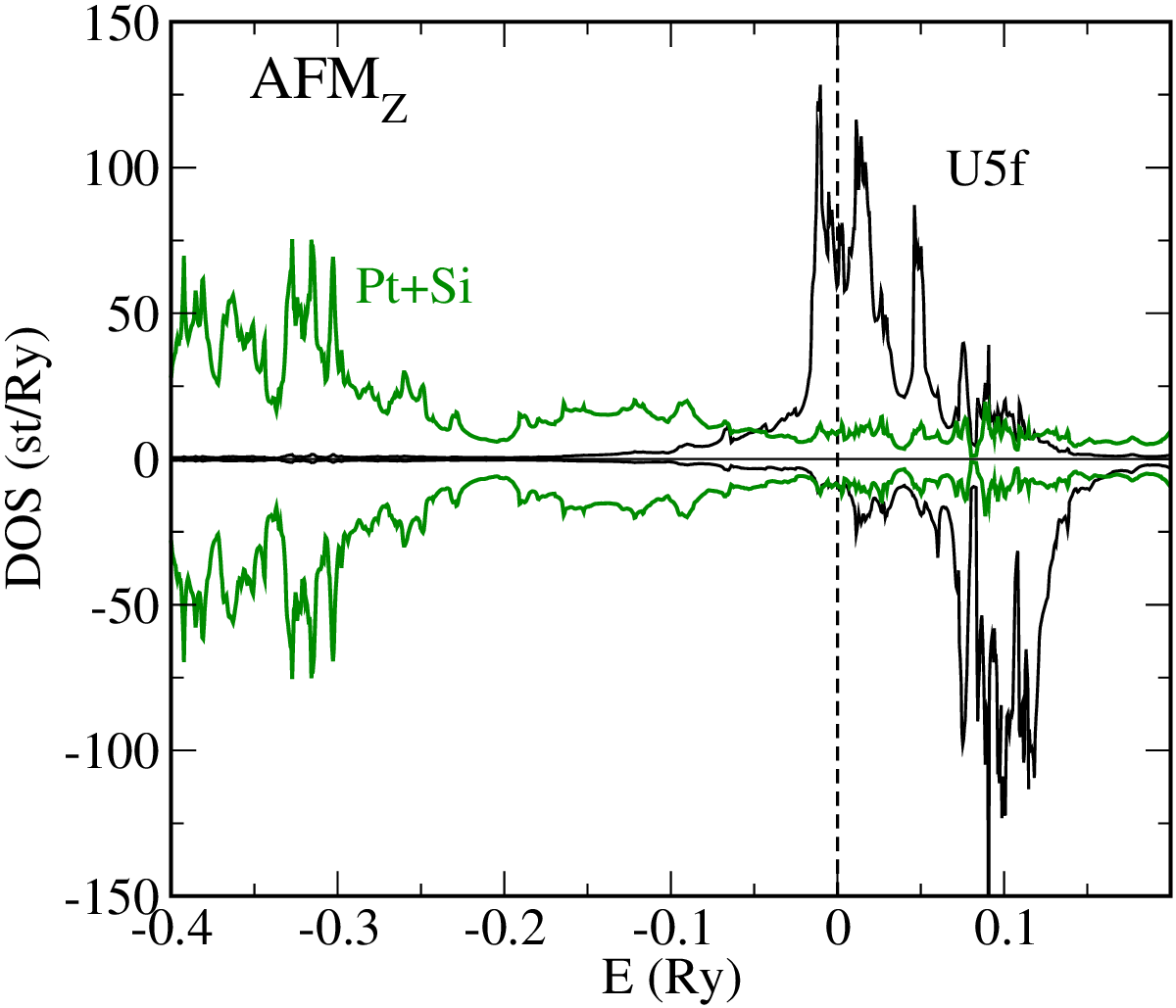}
\caption{The U 5f DOS and the sum of the Pt and Si DOSs.} 
\label{fig_DOSs_wideErange}
\end{figure}
In Fig.~\ref{fig_DOSs_wideErange} we show the U 5f DOS and 
the sum of the Pt and Si DOSs of the AFM$_Z$ configuration in a wide energy range. 
The common structure elements, i.e. common local maxima
and minima of the curves, reveal the hybridization of the 5f orbitals and delocalized
Pt and Si states. Also in the energy region where the partial DOS of the 5f orbitals
is very large, the admixture of the Pt and Si states is important.

\section{Comparison of mDOSs in GGA+$U$ and T5FEG calculations}
\label{sec_mDOS_T5FEG}
\begin{figure}[t]
\includegraphics*[width=8cm]{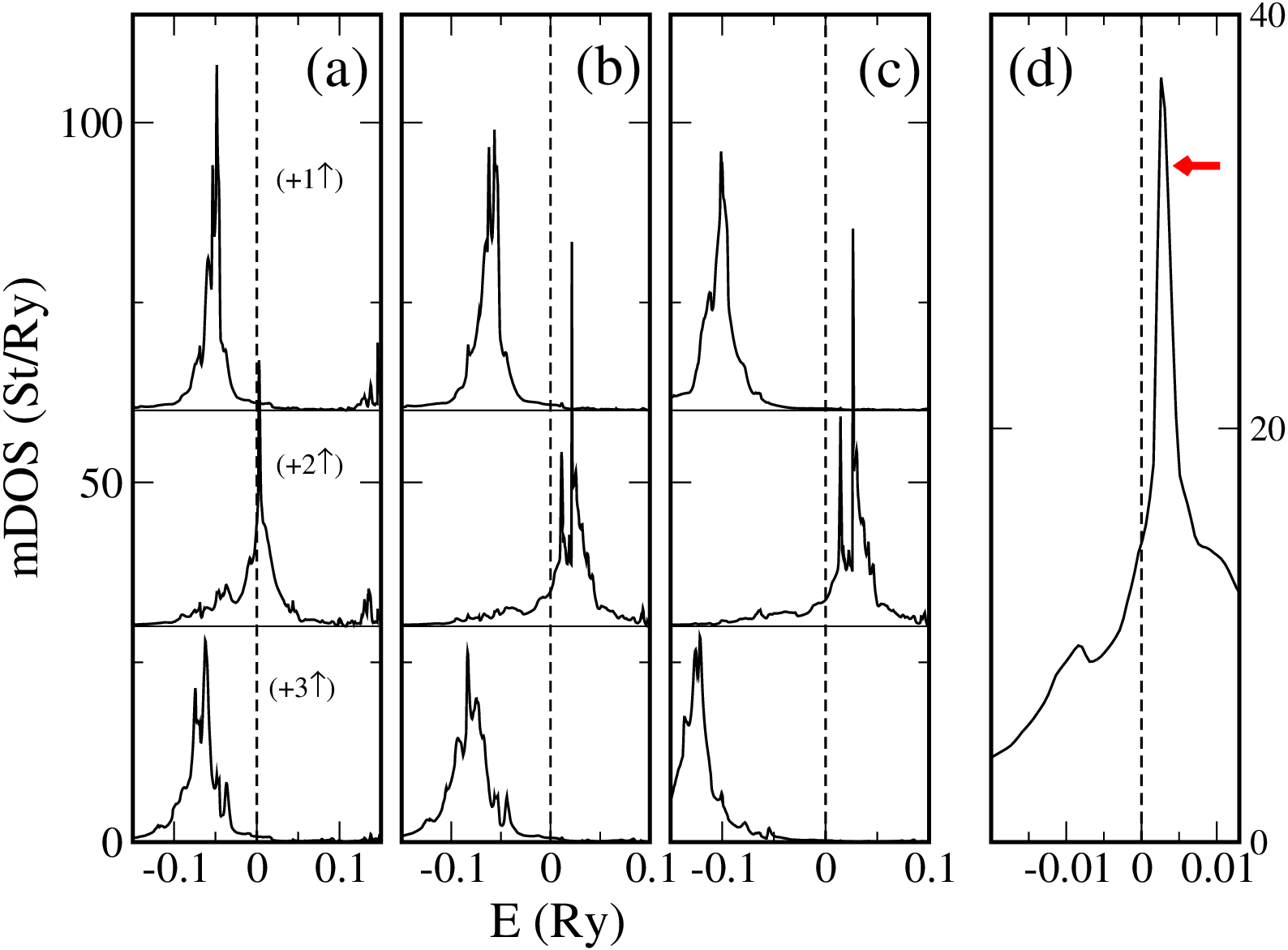}
\caption{Panels (a)-(c): Comparison of the low-energy mDOSs of the AFM$_Z$  configuration
calculated with the standard GGA+$U$ method and $U_H$=0.1~Ry (a), and using the 
GGA+T5FEG modification with $U_H$=0.1Ry (b) and 0.15Ry (c). Panel (d): zooming into the Fermi energy
region of the $m=2$ mDOS from panel (a). The arrow in panel (d) marks the mDOS peak that
may play important role in the formation of the Stoner continuum.
}
\label{fig_mDOS_T5FEG}
\end{figure}
In Fig.~\ref{fig_mDOS_T5FEG} we present the 
spin-majority mDOSs
for $m$ equal to $3$, $2$, and $1$ obtained in different calculations.
Figure~\ref{fig_mDOS_T5FEG}(a) shows the result of standard GGA+$U$ calculation
with $U_H$=0.1~Ry. Figures~\ref{fig_mDOS_T5FEG}(b),(c) present the results of the 
T5FEG-type modification of the method with $U_H$=0.1~Ry
and $U_H$=0.15~Ry, respectively. 

The comparison of the (a) and (b) panels 
shows that the application of $U_H$ to only $m=3$ and $m=1$ orbitals results in their deeper energy 
position. and somewhat 
higher occupations than in the standard case. The trend continues with increasing $U_H$, 
Fig.~\ref{fig_mDOS_T5FEG}(c). 
The position of the $m=2$ orbitals with respect to the Fermi level does not change importantly.
The corresponding occupation numbers are given in Fig.~\ref{fig_m_of_U} and discussed in the main text.


\begin{thebibliography}{99}
\bibitem{Fawcett1988}
An example of the system where the local moment picture is not applicable is fcc-Cr
where the magnetism is of the spin-density-wave type and caused by the 
nesting features of the Fermi surface: 
Eric Fawcett,
Rev. Mod. Phys. {\bf 60}, 209 (1988).
\bibitem{comment_electron_number}
In the ASW calculation,
the electron occupation numbers appear as a result of the mathematical 
procedure consisting of the evaluation of the contributions of certain
spherical harmonics in the decomposition of the crystal
wave functions in the U atomic area. The atom-centered basis
functions, augmented spherical waves, used in the method are
complex constructs treating differently the atomic region of
the central atom, the atomic regions of other atoms, and the
interstitial region. As a result, for the U atom there
are two different radial wave functions with l = 3 contributing
to the decomposition of the crystal wave functions: one corresponding 
to the central atom (Hankel-type function) and the other
obtained through the augmentation to the tails of the basis
functions centered at other atoms (Bessel-type function).  
In Fig.~\ref{FIG_occ__of_U}, we show the central-atom type contribution.
\end{thebibliography}
\end{document}